\documentstyle[preprint,aps,psfig]{revtex}
\begin{document}
\tightenlines
\draft

\title{Isotopic dependence of fusion cross sections
in reactions with heavy nuclei
}
\author{ G.G.Adamian$^{1,2,3}$, N.V.Antonenko$^{1,2}$,
W.Scheid$^{1}$
}
\address{$^{1}$Institut f\"ur Theoretische Physik der
Justus--Liebig--Universit\"at,
D--35392 Giessen, Germany\\
$^{2}$Joint Institute for Nuclear Research, 141980 Dubna, Russia\\
$^{3}$Institute of Nuclear Physics,
Tashkent 702132, Uzbekistan
}
\date{\today}
\maketitle

\begin{abstract}
The dependence of fusion cross section on the isotopic
composition of colliding nuclei is analysed
within the dinuclear system concept for compound nucleus formation.
Probabilities of fusion and surviving probabilities, ingredients of the
evaporation residue cross sections, depend decisively on the
neutron numbers of the dinuclear system.
Evaporation residue cross sections for the production
of actinides and superheavy nuclei, listed in table form,
are discussed and compared with existing experimental data.
Neutron-rich radioactive projectiles are shown to lead to
similar fusion cross sections as stable projectiles.
\end{abstract}


The synthesis of superheavy elements ($Z$=106-112) was reached by cold
fusion of heavy ions with lead and bismuth targets \cite{SHOF,Muen}.
Hot fusion reactions using $^{232}$Th, $^{238}$U and $^{242,244}$Pu targets
were also applied to synthesize the elements with
$Z$=110, 112 and 114 \cite{O1}.
A possible next step is to explore the synthesis
of heaviest nuclei with radioactive beams \cite{Muen,Hab,Muen1}.
Microscopical investigations of such planned experiments
are a challenge for theory. Usually the surviving probability
$W_{sur}$ of the formed compound nucleus
against fission in the de--excitation process is considered
as the crucial factor which is mainly responsible for
producing heavy and superheavy elements.
With neutron--reach projectiles one can obtain a
large stability (large $W_{sur}$) of the compound nucleus.
However, the probability of complete fusion $P_{CN}$,
dependent on nuclear structure effects and on the
neutron excess above the  nearest closed shells in the colliding nuclei,
is also very important for the correct calculation of the evaporation
residue cross section $\sigma_{ER}$.
For example, experimentally extracted probabilities $P_{CN}$
are strongly decreased \cite{SM} when the neutron numbers of the projectile
or target deviate from magic numbers.

The existing fusion models can be distinguished
by their choice of the relevant
collective degree of freedom responsible for complete
fusion. Many models assume an adiabatic
melting of the nuclei along the relative
distance $R$ of nuclear centers
(or the elongation of the system) \cite{P1,P2,Fr,Ro,Bar,P4}.
However, it was demonstrated
that the adiabatic scenario of fusion along the relative distance leads to
a large overestimation  and an incorrect isotopic trend of the
fusion probability \cite{AAIS}.
The dinuclear system (DNS) concept \cite{VV,AA,AA1,AA2,JNM,Cher0}
assumes that the united system is reached by a series of transfers
of nucleons or small clusters from the light nucleus to the heavier one
in a touching configuration. So, the dynamics of fusion is considered as a
diffusion of the DNS in the mass asymmetry, defined by
$\eta=(A_1-A_2)/(A_1+A_2)$ ($A_1$ and $A_2$ are the mass  numbers
the DNS nuclei), where the potential barrier
$B_{\rm fus}^*$ in $\eta$ supplies a  hindrance for fusion.

The assumption of a touching configuration of the two
reacting nuclei in the DNS model is supported by the structural
forbiddenness of fusion  \cite{Chu1,Chu2} which hinders
the nuclei to melt together along
the relative distance. This aspect is phenomenologically
described with a double folding potential in frozen density approximation
which shows a minimum near the touching distance of the nuclei \cite{Max}.
There are also experimental evidences \cite{To,Hind}
that the mass asymmetry degree of freedom
equilibrates more rapidly than the
elongation of the system. During the characteristic
time of fusion a statistical approach is
applicable to treat the evolution of the DNS which
also includes a diffusion to larger relative distances between
the centers of the nuclei
describing the quasi--fission process (decay of the DNS) competing with
the complete fusion.
In reactions with heavy nuclei the quasi--fission channel
dominates and leads to a strong reduction
of the fusion
\cite{AA,AA1,AA2,JNM,Cher0}.

In accordance with the DNS concept the evaporation
residue cross section is factorized as follows
\cite{AA2}
\begin{equation}
\sigma_{ER}(E_{\rm cm}) =
\sigma_c(E_{\rm cm})P_{CN}(E_{\rm cm},J=0)W_{sur}(E_{\rm cm},J=0).
\label{ER_eq}
\end{equation}
The calculations of the evaporation residue cross sections
demands an analysis of all three factors in (\ref{ER_eq}).
The value of $\sigma_c$ is the effective capture cross section
for the transition of the colliding
nuclei over the entrance (Coulomb) barrier with the probability $T$:
$$\sigma_c(E_{\rm cm}) = \pi\lambdabar^2(J_{max}+1)^2T(E_{\rm cm},J=0).$$
The contributing angular momenta in the evaporation residue
cross section are limited by the surviving probability
$W_{sur}(E_{\rm cm},J)$ with $J_{max}\approx 10-20$ when
highly fissile superheavy nuclei are produced \cite{PRL}. This
corresponds to almost central collisions with
impact parameters smaller than 1 fm.
The value of $J_{max}$ is  smaller than the critical angular momentum
$J_{crit}$ which restricts the capture cross section.
For reactions leading to superheavy nuclei, values of $J_{max}=10$
and $T(E_{\rm cm},J=0)=0.5$ are chosen for energies $E_{\rm cm}$ near the
Coulomb barrier. The capture cross sections obtained with these
parameters are in agreement with the ones calculated within a
microscopical model \cite{JNM}.

The probability of complete fusion $P_{CN}$ in (\ref{ER_eq})
depends on the competition between complete fusion
and quasi--fission after the capture stage. It can be expressed by
 rates in the quasi-stationary case as follows
\begin{eqnarray}
P_{CN}= 
\frac{\lambda_{\eta}^{Kr}}{\lambda_R^{Kr}+\lambda_\eta^{Kr}}-
\frac{\lambda_{\eta}^{Kr}\lambda_R^{Kr}}{\lambda_R^{Kr}+\lambda_\eta^{Kr}}
\frac{\tau_\eta-\tau_R}{1.72}.
\label{pcn_eq}
\end{eqnarray}
As in Ref.~\cite{AA1} we use a two-dimensional Kramers-type expression
(quasi--stationary solution of the Fokker-Planck equation)
with the quasi-stationary
rates of fusion $\lambda ^{Kr}_{\eta}$ and quasi--fission
$\lambda ^{Kr}_{R}$ through the fusion barrier ($B^*_{\rm fus}$)
in  $\eta$ and quasi--fission barrier ($B_{\rm qf}$) in  $R$,
respectively.
The second term in (\ref{pcn_eq}) is related
to the transient times $\tau_R$ and $\tau_\eta$ to reach
the quasi--stationary rates along the $R$ and $\eta$ coordinates
($\tau^{-1}_R,\tau^{-1}_{\eta}>\lambda ^{Kr}_{R},\lambda ^{Kr}_{\eta}$)
\cite{AA1}.

In the case that the fusion barrier is much
higher than the quasi-fission barrier,
$B^*_{\rm fus} \gg B_{\rm qf}$, i.e. if the transient time
$\tau_{\eta}$ in $\eta$ is larger (or equal) than the lifetime $t_0$
of the initial DNS, we obtain \cite{AA1}
\begin{eqnarray}
P_{CN}=\frac{\lambda_{\eta}^{Kr}}{1.72}
[\tau_\eta(\exp[t_0/\tau_\eta]-1)-t_0].
\label{pcn1_eq}
\end{eqnarray}
Since the pocket in the nucleus-nucleus potential becomes
very shallow ($B_{\rm qf}\approx 0$)
in reactions with large $Z_1\times Z_2$,
the lifetime $t_0$ of the DNS is strongly depressed with increasing
bombarding energy $E_{\rm c.m.}$
above the Coulomb barrier.
Due to this,  the value of $P_{CN}$ in Eq.(3) is smaller than the one
in the quasi--stationary regime, given
by Eq.~(2), which can not be reached in this case \cite{AA1}.

The surviving probability under the evaporation of $x$ neutrons
is considered according to \cite{AA2,Cher} as
\begin{equation}
W_{sur}(E_{CN}^*,J)\approx P_{xn}(E_{CN}^*,J)
\prod\limits_{i=1}^x
\frac{\Gamma_n(E_{CN_i}^*,J_i)}
{\Gamma_n(E_{CN_i}^*,J_i)+\Gamma_f(E_{CN_i}^*,J_i)}.
\label{wsur_eq}
\end{equation}
Here, $P_{xn}$ is the probability
of realization of the $xn$ channel
at the excitation energy $E_{CN}^*$
of the compound nucleus, $i$ the index of
evaporation step, $\Gamma_n$ and $\Gamma_f$
are the partial widths of neutron emission and fission.
$E_{CN_i}^*$ and $J_i$ are the mean values of excitation energy
and angular momentum quantum number of the compound nucleus,
respectively, at the beginning of step $i$ with
$E_{CN_1}^*=E_{CN}^*$ and $J_1=J$.
In the calculation of $W_{sur}$ we used the microscopic corrections
of M\"oller and Nix \cite{Moel} as fission barriers. The neutron binding
energies are also taken from \cite{Moel}.

The barriers $B^*_{\rm fus}$
and $B_{\rm qf}$ are given by the potential energy of the DNS which
is calculated as the sum of binding energies $B_i$ of the nuclei (i=1,2) and
of the nucleus-nucleus potential $V(R,\eta,J)$ \cite{AA,AA1,AA2}:
\begin{eqnarray}
U(R, \eta, J)&= &B_1+ B_2+  V(R,\eta,J)- [B_{12}+V^{'}_{rot}(J)].
\label{Pot_eq}
\end{eqnarray}
The shell effects are included in the binding energies. The isotopic
composition of the nuclei forming the DNS is obtained with the condition
of a $N/Z$-equilibrium in the system.
The value of $U(R, \eta, J)$ is normalized to the energy
of the rotating compound nucleus by $B_{12}+V^{'}_{rot}$.
Deformation effects are taken into account in the calculation
of the potential energy surface \cite{AA2}. The heavy nuclei
in the DNS, which are deformed in the ground state, are treated
with the parameters
of deformation taken from Ref.~\cite{Rama}.
The light nuclei of the DNS are assumed to be deformed only
if the energy of their $2^+$ state is smaller than 1.5 MeV.
As known from experiments on sub--barrier
fusion of lighter nuclei, these states are easily populated.
For the collision energies considered here,
the relative orientation of the nuclei in the DNS follows
the minimum of the potential energy during the evolution in $\eta$.

The experimentally observed hindrance of the fusion roughly
increases with  growing Coulomb repulsion between the
colliding nuclei, but also their
shell structure and isotopic composition play a major role
\cite{SM,QUI,SAH,KRXE}. In Table~1 we present calculated excess energies
above the entrance Coulomb barrier in the DNS model for various
reactions and compare them with the surplus of energy
extracted from experimental data above the corresponding Bass
barriers \cite{BERD}. In these calculations we did not average
the inner fusion barrier $B^*_{\rm fus}$
over all possible orientations of colliding nuclei
as we usually do in the calculations of $P_{CN}$ and
$\sigma_{ER}$, taking approximately the half of the
deformation parameters of the nuclei of the entrance channel.
The obtained energy thresholds in $\eta$ are maximal ones.
They are not always in good
agreement with the data extracted
from experiment because these data are
not directly measurable but are
obtained with model assumptions about $P_{CN}$ and $W_{sur}$.

As shown in Table~1 the isotopic trends of the DNS
model agree with the experimental ones. The
energy thresholds for fusion increase and, correspondingly, the
fusion probabilities decrease \cite{QUI,SAH}
when the neutron number of projectile
or target deviates more from a magic number
in the reactions  $^{90}$Zr+$^{90,92,96}$Zr, $^{90,96}$Zr+$^{100}$Mo
$^{86}$Kr+$^{99,102,104}$Ru, $^{90,92,94,96}$Zr+$^{124}$Sn and
$^{86}$Kr+$^{130,136}$Xe.
This effect is simply explained by the deformation of the nuclei
in the initial DNS and DNS at the top
of the barrier in $\eta$ and by the shell effects in dependence of the DNS
potential energy on $\eta$.
For example, the value of the energy threshold
for fusion, which determines the fusion probability,
is larger in the $^{86}$Kr+$^{130}$Xe reaction
than in the $^{86}$Kr+$^{136}$Xe reaction \cite{KRXE}.
Since in addition the surviving probability $W_{sur}$
is larger in the reaction with
$^{136}$Xe  than in the reaction with $^{130}$Xe, there results an
experimental difference of
about 3 orders of magnitude in $\sigma_{ER}$ in these reactions \cite{KRXE}.
For most reactions, for example,
$^{90}$Zr+$^{90}$Zr, $^{100}$Mo+$^{100}$Mo and
$^{110}$Pd+$^{110}$Pd, we obtained evaporation residue
cross sections with the values of  $P_{CN}$ of the DNS model which are
in good agreement with the experimental
cross sections \cite{AA,AA1}. In contrast, models which
treat fusion as a motion in $R$, give an incorrect isotopic trend
of $P_{CN}$. In these models $P_{CN}$ always increases
with the neutron number above the nearest closed shell
\cite{SM,BERD} because an increasing deformation of the nuclei
in the entrance channel effectively lowers the barrier.

Since the evaporation residue cross section increases with
the number of neutrons in all reactions listed in Table~1,
the value of $W_{sur}$ has to
grow faster than $P_{CN}$ decreases.
In fusion reactions leading to actinides, for example in the
$^{66,76}$Zn+$^{174}$Yb reaction, the increase
of $W_{sur}$ with the neutron number of the system
is stronger than the decrease of $P_{CN}$
This effect, shown in Fig.~1 for reactions $^{A}$Zn+$^{174}$Yb, gives
a certain preference for
neutron-rich projectiles in
producing actinides. Note that the numbers of neutrons
in the nuclei $^{66}$Zn and $^{76}$Zn are close to different magic numbers.

Fusion probabilities in symmetric and almost
symmetric reactions with heavy nuclei like $^{124,132}$Sn and $^{136,142}$Xe
strongly depend on the model of fusion.
For example, in an adiabatic treatment, where the fusion
is mainly described by the dynamics in
the relative distance coordinate with an increasing neck,
we found $P_{CN}\approx 10^{-6}$ and $\sigma_{ER}\approx$30pb
for the reaction $^{132}$Sn + $^{132}$Sn$\to ^{261}$Fm+$3n$.
In the DNS model the values of $P_{CN}$ and, correspondingly,
$\sigma_{ER}$ are about three orders of magnitude smaller.
According to the DNS concept, cross sections for the
synthesis of the heaviest elements in nearly symmetric
reactions are very small due to
small fusion probabilities, for example, in the reactions
$^{136}$Xe + $^{136}$Xe$\to ^{272}$Hs,
$^{142}$Xe + $^{150,154}$Nd$\to ^{292,296}$114,
$^{132}$Sn + $^{160}$Gd$\to ^{292}$114 and
$^{137}$Te + $^{158}$Sm$\to ^{295}$114.
Experimental data on symmetric reactions with stable and
radioactive beams could help to prove the DNS model
for the fusion process and would give information about the
time for the transition from the diabatic to adiabatic regime
(the time of suppression of the structure forbiddenness
for melting of nuclei \cite{Chu1}).

In contrast to other models, the optimal excitation energy
$E^*_{CN}$ of the compound nucleus and evaporation residue cross section
$\sigma_{ER}$ in cold fusion reactions with
stable projectiles are reproduced in the DNS concept
\cite{AA2}. These results are listed in Table~2 for reactions
leading to the Fm element and Pb- and Bi-based reactions.
The evaporation residue cross sections are
compared with the experimental data of Refs.~\cite{SHOF,Gagg}.
All other cross sections are predictions of the present version
of the DNS model. Figs.~2a) and 2b) show the fusion probabilities
and the optimal excitation energies of the compound nuclei,
respectively, for $^{208}$Pb, $^{209}$Bi$(^AX,1n)$ reactions.

The values of the optimal excitation energy $E^*_{CN}$ are calculated
by applying theoretical $Q$-values of Refs.~\cite{Moel}.
They increase for $Z>112$. $Q$-values of Ref.~\cite{Moel1}
are slightly different for $Z>113$.
As in the case of reactions with heavy nuclei mentioned above,
the calculated values of $P_{CN}$ are maximal when
the neutron number of the projectile is
equal to a magic number, for example, in the reactions
$^{82}$Ge+$^{208}$Pb, $^{84}$Se+$^{208}$Pb and $^{86}$Kr+$^{208}$Pb.
The decrease
of the cold fusion cross section by four orders of magnitude
from $Z=$104 to 112 is mainly caused by a decrease of
$P_{CN}$  due to a strong competition between
complete fusion and quasi--fission in the DNS (see Fig.~2a)).
For the  reaction $^{70}$Zn+$^{208}$Pb$\to ^{277}$112+$1n$,
 $\sigma_{ER}\approx $1pb  is practically on the level of the present
experimental possibilities.
In reactions $^{74,76}$Ge+$^{208}$Pb$\to ^{283,281}$114+$1n$
we expect a value of $\sigma_{ER}$  which is smaller than 0.2 pb.
The values of $\sigma_{ER}$  for the  $Z$=116 and 118 elements formed
in the $^{84}$Se, $^{86}$Kr+$^{208}$Pb reactions are
of the order of 0.01 pb (Table~2).

The values of $W_{sur}$ in Table~2 were calculated with
the theoretical data of Ref.~\cite{Moel}. One can see that
 characteristic values of $W_{sur}$ are about
$10^{-3}-10^{-4}$ for nuclei with $Z$=104--113 and  about $10^{-2}$
for nuclei with $Z$=114, 116 and 118.
The proton magic number 114 in the region
of the stability island \cite{Moel,Moel1,Sobi}
leads to a larger increase of $W_{sur}$.
The large $W_{sur}$ of the nuclei $^{292}$114, $^{294}$116 and
$^{296}$118 arises due to the fact that
the neutron number in these nuclei is equal to the theoretically
predicted magic number $N$=178 \cite{Moel,Moel1}.
When the number of neutrons deviates from this
magic number, $W_{sur}$ decreases.
The surviving probabilities in the reactions
$^{70}$Zn+$^{208}$Pb and $^{74}$Ge+$^{208}$Pb were
calculated with the data of Ref.~\cite{Moel1}
because $W_{sur}$ becomes unrealistically small
with the data of Ref.~\cite{Moel} (about two orders
of magnitude in comparison to the neighbouring nuclei).
Using the microscopical corrections and neutron
binding energies from Ref.~\cite{Moel1} instead of Ref.~\cite{Moel}
we obtained even smaller cross sections $\sigma_{ER}$ of
50, 5 and 0.8 fb for the reactions with the projectiles
$^{76}$Ge, $^{82}$Se and $^{86}$Kr on $^{208}$Pb,
respectively.
With the fission characteristics of Ref.~\cite{Sobi}
we get evaporation residue cross sections
$\sigma_{ER}$ which are again smaller than the
latter ones. In conclusion the cross sections in Table~2 are
optimistic estimates.

Let us consider whether  the expected values
of evaporation residue cross sections are larger
with  radioactive projectiles.
In the Pb-based reactions with neutron-rich nuclei
$^{70,74,78}$Ni, $^{80}$Zn, $^{78-86}$Ge, $^{84-92}$Se and $^{88-92}$Kr
the inner fusion barrier $B^*_{fus}$
in mass asymmetry varies between 12 and 22
MeV. In order to overcome this barrier, the initial DNS must have
excitation energies which lead to an excited compound nucleus
with the possibility of an 1$n$ or 2$n$ emission.
The calculated cross sections for some possible reactions are presented in
Table~2. Bombarding energies near the Coulomb barrier
lead to maximal evaporation residue cross sections.
In these reactions the increase of $W_{sur}$ is compensated by a
decreasing fusion probability $P_{CN}$ and the value of $\sigma_{ER}$
depends weakly on the isotopic  composition of the colliding nuclei.
The values of $P_{CN}$ and $W_{sur}$
for the reactions $^{A}$Ni+$^{208}$Pb and $^{A}$Ge+$^{208}$Pb
are presented in Fig.~3 as functions
of $A$. The calculations were performed with the same parameters
as used for the stable projectiles and are in good agreement with
experimental data (Table~2) \cite{SHOF,AA2}.
Due to deformation effects and binding energies
of the nuclei in the DNS, the dependence of $P_{CN}$ on $A$
can have some minima and maxima.

The yield of the element $Z=110$ results larger
in the $^{78}$Ni+$^{208}$Pb reaction than in the
$^{62,64}$Ni+$^{208}$Pb reactions.
In the $^{70,74}$Ni+$^{208}$Pb reactions the values of $\sigma_{1n}$
are close to the experimental value of
$\sigma_{1n}=3.5^{+2.7}_{-1.8}$ pb of the
$^{62}$Ni+$^{208}$Pb reaction. In the $^{70,74}$Ni+$^{208}$Pb reactions
the cross sections $\sigma_{2n}$ are about 4 times smaller than $\sigma_{1n}$
due to smaller values of $\sigma_c$ and $W_{sur}$.
In spite of the large values of $W_{sur}$ in the reactions
$^{84,86}$Ge+$^{208}$Pb, $^{86,88,90,92}$Se+$^{208}$Pb
and $^{88,90,92}$Kr+$^{208}$Pb, the corresponding
values of $\sigma_{ER}$ are expected
to be smaller than 0.1 pb due to the very small values of
$P_{CN}$ (Table~2 and Fig.~3).

In spite of the expected relatively small yields
for neutron--rich superheavies
the larger lifetime of these nuclei will allow a detailed study
of their properties. The
lifetime of molecular--type configurations with an initial DNS
in the entrance channel can be studied with beams of radioactive
nuclei. In reactions with neutron--rich nuclei, a neutron emission
can occur out of the DNS besides a possible quasi-fission
because the characteristic emission time
becomes comparable with
the fusion time. This process decreases the excitation energy
of the DNS and the fusion probability.
The effect of neutron emission from the
DNS is expected to be important for energies larger than the energy in the
3$n$ channel. With a neutron emission from the DNS the fusion process
is more complex and has to be studied.

Intensive beams of neutron-rich nuclei are very useful
for producing heavy actinides, e.g. Fm as listed in Table~2.
In the Pb-based reactions the use
of neutron-rich projectiles leads to values of $\sigma_{ER}$
comparable with evaporation residue cross sections for
reactions with the stable projectiles. More asymmetric
reactions with radioactive beams could be more useful in the
production of superheavies.

\vspace*{1cm}
We thank Profs. D.Habs, R.V.Jolos, G.M\"unzenberg, V.V.Volkov,
Drs. E.A.Cherepanov, S.P.Ivanova and A.K.Nasirov for fruitful
discussions and suggestions. G.G.A. is grateful to Alexander
von Humboldt-Stiftung (Bonn) for support.
This work was supported in part by DFG and RFBR.

\begin{table}[here]
\caption{Calculated maximal energy excess $\Delta E^{th}$
above the entrance Coulomb barrier in the DNS model and
the surplus of
energy $\Delta B^{exp}$ above the Bass barrier
extracted from experimental data \protect\cite{BERD}.
The number of valence neutron particles or holes to the
nearest closed shell in the projectile or target is
denoted by $\Delta N$.}
\begin{tabular}{|c|c|c|c|c|c|c|c|}
Reactions & $\Delta N$ & $\Delta E^{th}$ & $\Delta B^{exp}$  &
Reactions & $\Delta N$ & $\Delta E^{th}$ & $\Delta B^{exp}$     \\
          &            &     (MeV)       &        (MeV)      &
          &            &     (MeV)       &        (MeV)         \\
\hline
$^{86}$Kr+$^{92}$Mo$\to^{178}$Pt   & 0 & 0.0  & $1.4^{+1.0}_{-1.0}$&
$^{98}$Mo+$^{100}$Mo$\to^{198}$Po  & 6 & 12.6 & $14.1^{+1.0}_{-1.0}$ \\
$^{86}$Kr+$^{100}$Mo$\to^{186}$Pt  & 8 & 2.0  & $4.3^{+2.0}_{-2.0}$&
$^{100}$Mo+$^{100}$Mo$\to^{200}$Po & 8 & 10.3 & $12.2^{+0.5}_{-0.5}$ \\
$^{86}$Kr+$^{99}$Ru$\to^{185}$Hg   & 5 & 3.6  & $3.1^{+1.2}_{-1.2}$&
$^{100}$Mo+$^{104}$Ru$\to^{204}$Rn & 10& 12.7 & $23.0^{+1.1}_{-1.1}$ \\
$^{86}$Kr+$^{102}$Ru$\to^{188}$Hg  & 8 & 5.2  & $6.5^{+1.3}_{-1.3}$&
$^{100}$Mo+$^{110}$Pd$\to^{210}$Ra & 14& 13.7 & $29.0^{+1.2}_{-1.2}$ \\
$^{86}$Kr+$^{104}$Ru$\to^{190}$Hg  & 10& 5.8 & $7.2^{+1.3}_{-1.3}$&
$^{90}$Zr+$^{124}$Sn$\to^{214}$Th  & 0 & 6.1 & $20.3^{+4.0}_{-4.0}$ \\
$^{90}$Zr+$^{90}$Zr$\to^{180}$Hg   & 0 & 2.9 & $0.0^{+0.5}_{-0.5}$&
$^{92}$Zr+$^{124}$Sn$\to^{216}$Th  & 2 & 6.6 & $20.8^{+4.0}_{-3.0}$ \\
$^{90}$Zr+$^{92}$Zr$\to^{182}$Hg   & 2 & 4.0 & $4.2^{+0.5}_{-0.5}$&
$^{94}$Zr+$^{124}$Sn$\to^{118}$Th  & 4 & 8.8 & $22.7^{+5.0}_{-3.0}$\\
$^{90}$Zr+$^{96}$Zr$\to^{186}$Hg   & 6 & 4.6 & $5.1^{+0.5}_{-0.5}$&
$^{96}$Zr+$^{124}$Sn$\to^{220}$Th  & 6 & 12.5 & $26.7^{+5.0}_{-3.0}$ \\
$^{96}$Zr+$^{96}$Zr$\to^{192}$Hg   & 6 & 6.9  & $4.2^{+1.2}_{-1.2}$&
$^{86}$Kr+$^{130}$Xe$\to^{116}$Th  & 6 & 7.8  & \\
$^{90}$Zr+$^{100}$Mo$\to^{190}$Pb  & 0 & 5.5  & $5.1^{+1.0}_{-1.0}$&
$^{86}$Kr+$^{136}$Xe$\to^{222}$Th  & 0 & 5.0  & \\
$^{92}$Zr+$^{100}$Mo$\to^{192}$Pb  & 2 & 7.0  & $5.8^{+1.0}_{-1.0}$&
$^{110}$Pd+$^{110}$Pd$\to^{220}$U  & 14& 20.9 &  \\
$^{96}$Zr+$^{100}$Mo$\to^{196}$Pb  & 6 & 8.8  & $9.5^{+1.0}_{-1.0}$&
$^{124}$Sn+$^{124}$Sn$\to^{248}$Fm & 8 & 23.2 &  \\
$^{92}$Mo+$^{100}$Mo$\to^{192}$Po  & 0 & 11.8 & $13.0^{+2.0}_{-2.0}$&
$^{132}$Sn+$^{132}$Sn$\to^{264}$Fm & 0 & 30.7 &  \\
$^{94}$Mo+$^{100}$Mo$\to^{194}$Po  & 2 & 14.9 & $16.3^{+1.0}_{-1.0}$&
$^{130}$Xe+$^{130}$Xe$\to^{260}$Hs & 6 & 37.7 &  \\
$^{96}$Mo+$^{100}$Mo$\to^{196}$Po  & 4 & 8.2  & $10.4^{+1.0}_{-1.0}$&
$^{136}$Xe+$^{136}$Xe$\to^{272}$Hs & 0 & 33.1 &  \\
\end{tabular}
\end{table}

\begin{table}[here]
\caption{Excitation energy $E_{CN}^*$ of compound nucleus,
fusion probability $P_{CN}$, capture cross section $\sigma_c$,
surviving probability $W_{sur}$, and theoretical
$\sigma_{ER}^{th}$  and experimental $\sigma_{ER}^{exp}$
evaporation residue cross sections for reactions leading
to the Fm nucleus and Pb-based reactions.
The experimental data are taken from Refs.\protect\cite{Gagg} and
\protect\cite{SHOF} (cold fusion).}
\begin{tabular}{|c|c|c|c|c|c|c|}
Reactions &  $E^*_{CN}$ & $P_{CN}$ & $\sigma_c$ & $W_{sur}$ & $\sigma_{ER}^{th}$
& $\sigma_{ER}^{exp}$\\
          &    (MeV)    &          &     (mb)   &           &
&               \\
\hline
$^{66}$Zn+$^{174}$Yb$\to^{238}$Fm+$2n$ & 26.0 & 4$\times 10^{-2}$ &
9.6  & 8.0$\times 10^{-7}$ & 0.3 nb &\\
$^{76}$Zn+$^{174}$Yb$\to^{248}$Fm+$2n$ & 23.0 & 2$\times 10^{-3}$ &
8.8  & 6.0$\times 10^{-4}$ & 10.6 nb &\\
$^{76}$Ge+$^{170}$Er$\to^{244}$Fm+$2n$ & 24.6 & 5$\times 10^{-4}$ &
8.4  & 3.0$\times 10^{-4}$ & 1.3 nb & $1.6^{+1.3}_{-1.6}$ nb \\
\hline
$^{50}$Ti+$^{208}$Pb$\to^{257}104+1n$  & 16.1 & 3$\times 10^{-2}$ &
5.3  & 9$\times 10^{-5}$ & 14.3 nb &$10^{+1.3}_{-1.3}$ nb\\
$^{50}$Ti+$^{209}$Bi$\to^{258}105+1n$  & 16.2 & 3$\times 10^{-3}$ &
5.2  & 3$\times 10^{-4}$ & 4.7 nb &$4^{+1.3}_{-1.6}$ nb\\
$^{54}$Cr+$^{208}$Pb$\to^{261}106+1n$  & 16.0 & 9$\times 10^{-4}$ &
4.6  & 1$\times 10^{-4}$ & 0.4 nb &$0.5^{+0.14}_{-0.14}$ nb\\
$^{54}$Cr+$^{209}$Bi$\to^{262}107+1n$  & 15.9 & 2$\times 10^{-4}$ &
4.5  & 3$\times 10^{-4}$ & 270 pb &$ 163^{+34}_{-34}$ pb\\
$^{58}$Fe+$^{208}$Pb$\to^{265}108+1n$  & 15.5 & 3$\times 10^{-5}$ &
4.0  & 4$\times 10^{-4}$ & 48 pb &$65.8^{+7.5}_{-7.5}$ pb\\
$^{58}$Fe+$^{209}$Bi$\to^{266}109+1n$  & 15.7 & 6$\times 10^{-6}$ &
4.0  & 5$\times 10^{-4}$ & 12 pb &$8.8^{+3.3}_{-3.3}$ pb\\
$^{62}$Ni+$^{208}$Pb$\to^{269}110+1n$  & 12.3 & 4.5$\times 10^{-6}$ &
3.5  & 5$\times 10^{-4}$ & 7 pb &$3.5^{+2.7}_{-1.8}$ pb\\
$^{64}$Ni+$^{208}$Pb$\to^{271}110+1n$  & 10.7 & 1$\times 10^{-5}$ &
3.4  & 5$\times 10^{-4}$ & 17 pb &$15^{+9}_{-6}$ pb\\
$^{70}$Ni+$^{208}$Pb$\to^{277}110+1n$  & 13.5 & 7$\times 10^{-8}$ &
3.1  & 5$\times 10^{-3}$ & 1.1 pb &\\
$^{74}$Ni+$^{208}$Pb$\to^{281}110+1n$  & 15.0 & 6$\times 10^{-8}$ &
3.0  & 2$\times 10^{-2}$ & 3.6 pb &\\
$^{78}$Ni+$^{208}$Pb$\to^{284}110+2n$  & 17.5 & 2$\times 10^{-7}$ &
3.0  & 6$\times 10^{-2}$ & 36 pb&\\
$^{64}$Ni+$^{209}$Bi$\to^{272}111+1n$  & 10.5 & 2$\times 10^{-6}$ &
3.4  & 6$\times 10^{-4}$ & 4.1 pb &$3.5^{+4.6}_{-2.3}$ pb\\
$^{70}$Zn+$^{208}$Pb$\to^{277}112+1n$  &9.8 & 1$\times 10^{-6}$ &
3.0  & 6$\times 10^{-4}$ & 1.8 pb &$1.0^{+1.3}_{-0.7}$ pb\\
$^{80}$Zn+$^{208}$Pb$\to^{286}112+2n$  & 15.7 & 7$\times 10^{-9}$ &
2.6  & 1$\times 10^{-1}$ & 1.8 pb& \\
$^{68}$Zn+$^{209}$Bi$\to^{276}113+1n$  & 9.6 & 1$\times 10^{-6}$ &
2.9  & 1$\times 10^{-4}$ & 290 fb &\\
$^{70}$Zn+$^{209}$Bi$\to^{278}113+1n$  & 10.6 & 4$\times 10^{-7}$ &
2.9  & 2$\times 10^{-4}$ & 232 fb & $<$600 fb\\
$^{74}$Ge+$^{208}$Pb$\to^{281}114+1n$  & 12.5 & 2$\times 10^{-8}$ &
2.5  & 2$\times 10^{-3}$ & 100 fb &\\
$^{76}$Ge+$^{208}$Pb$\to^{283}114+1n$  & 12.4 & 4$\times 10^{-9}$ &
2.5  & 2$\times 10^{-2}$ & 200 fb &\\
$^{78}$Ge+$^{208}$Pb$\to^{285}114+1n$  & 14.2 & 5$\times 10^{-10}$ &
2.1  & 2$\times 10^{-2}$ & 21 fb &\\
$^{82}$Ge+$^{208}$Pb$\to^{289}114+2n$  & 16.3 & 1$\times 10^{-9}$ &
2.0  & 1$\times 10^{-1}$ & 200 fb &\\
$^{84}$Ge+$^{208}$Pb$\to^{291}114+2n$  & 18.5 & 2$\times 10^{-10}$ &
2.0  & 2$\times 10^{-1}$ & 80 fb &\\
$^{86}$Ge+$^{208}$Pb$\to^{294}114+2n$  & 20.4 & 4$\times 10^{-10}$ &
2.0  & 4$\times 10^{-2}$ & 32 fb &\\
$^{82}$Se+$^{208}$Pb$\to^{289}116+1n$  & 13.8 & 4$\times 10^{-10}$ &
1.9  & 2$\times 10^{-2}$ & 15 fb &\\
$^{84}$Se+$^{208}$Pb$\to^{291}116+1n$  & 14.6 & 7$\times 10^{-10}$ &
1.8  & 2$\times 10^{-2}$ & 25 fb &\\
$^{86}$Se+$^{208}$Pb$\to^{293}116+2n$  & 14.8 & 1$\times 10^{-10}$ &
1.8  & 6$\times 10^{-2}$ & 11 fb &\\
$^{88}$Se+$^{208}$Pb$\to^{295}116+2n$  & 15.0 & 8$\times 10^{-11}$ &
1.8  & 2$\times 10^{-2}$ & 2.9 fb &\\
$^{90}$Se+$^{208}$Pb$\to^{297}116+2n$  & 14.8 & 1$\times 10^{-10}$ &
1.8  & 2$\times 10^{-2}$ & 3.6 fb &\\
$^{92}$Se+$^{208}$Pb$\to^{299}116+2n$  & 20.2 & 1.5$\times 10^{-10}$ &
1.8  & 6$\times 10^{-3}$ & 1.7 fb &\\
$^{84}$Kr+$^{208}$Pb$\to^{291}118+1n$  & 12.5 & 5$\times 10^{-11}$ &
1.7  & 2$\times 10^{-2}$ & 1.7 fb &\\
$^{86}$Kr+$^{208}$Pb$\to^{293}118+1n$  & 13.3 & 1.5$\times 10^{-10}$ &
1.7  & 2$\times 10^{-2}$ & 5.1 fb &\\
$^{88}$Kr+$^{208}$Pb$\to^{295}118+1n$  & 12.0 & 3$\times 10^{-11}$ &
1.7  & 8$\times 10^{-2}$ & 4.1 fb &\\
$^{90}$Kr+$^{208}$Pb$\to^{297}118+1n$  & 13.1 & 1.5$\times 10^{-11}$ &
1.6  & 5$\times 10^{-2}$ & 1.2 fb &\\
$^{92}$Kr+$^{208}$Pb$\to^{299}118+1n$  & 12.4 & 1.5$\times 10^{-11}$ &
1.6  & 4$\times 10^{-2}$ & 1.0 fb &\\
\end{tabular}
\end{table}

\begin{figure}
\psfig{figure=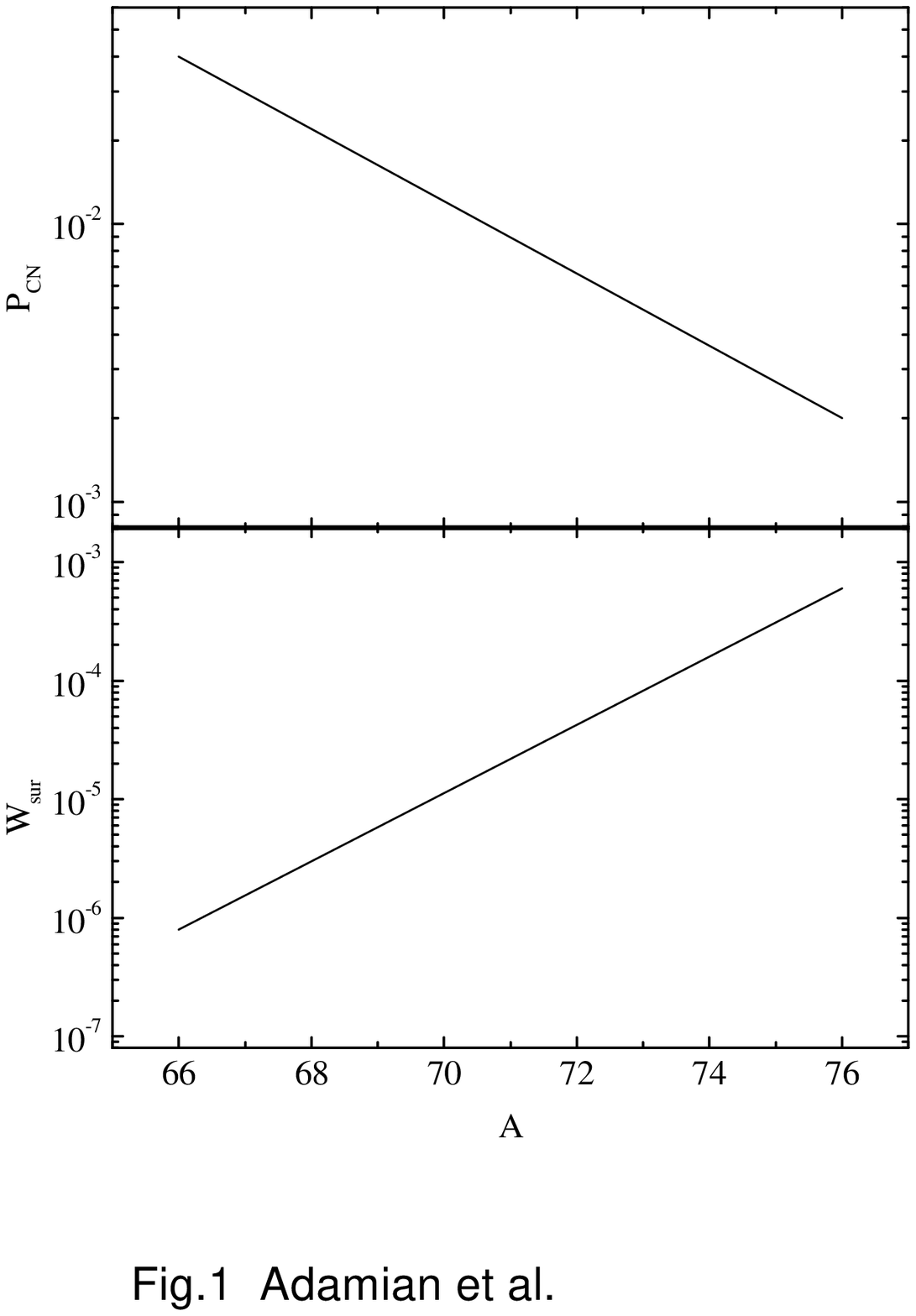,width=14cm}
\caption{Fusion ($P_{CN}$) and surviving ($W_{sur}$)
probabilities as functions of mass number $A$
of the projectile in reactions $^{A}$Zn+$^{174}$Yb at bombarding
energies supplying the maximum yield of evaporation residues.}
\label{1_fig}
\end{figure}

\begin{figure}
\psfig{figure=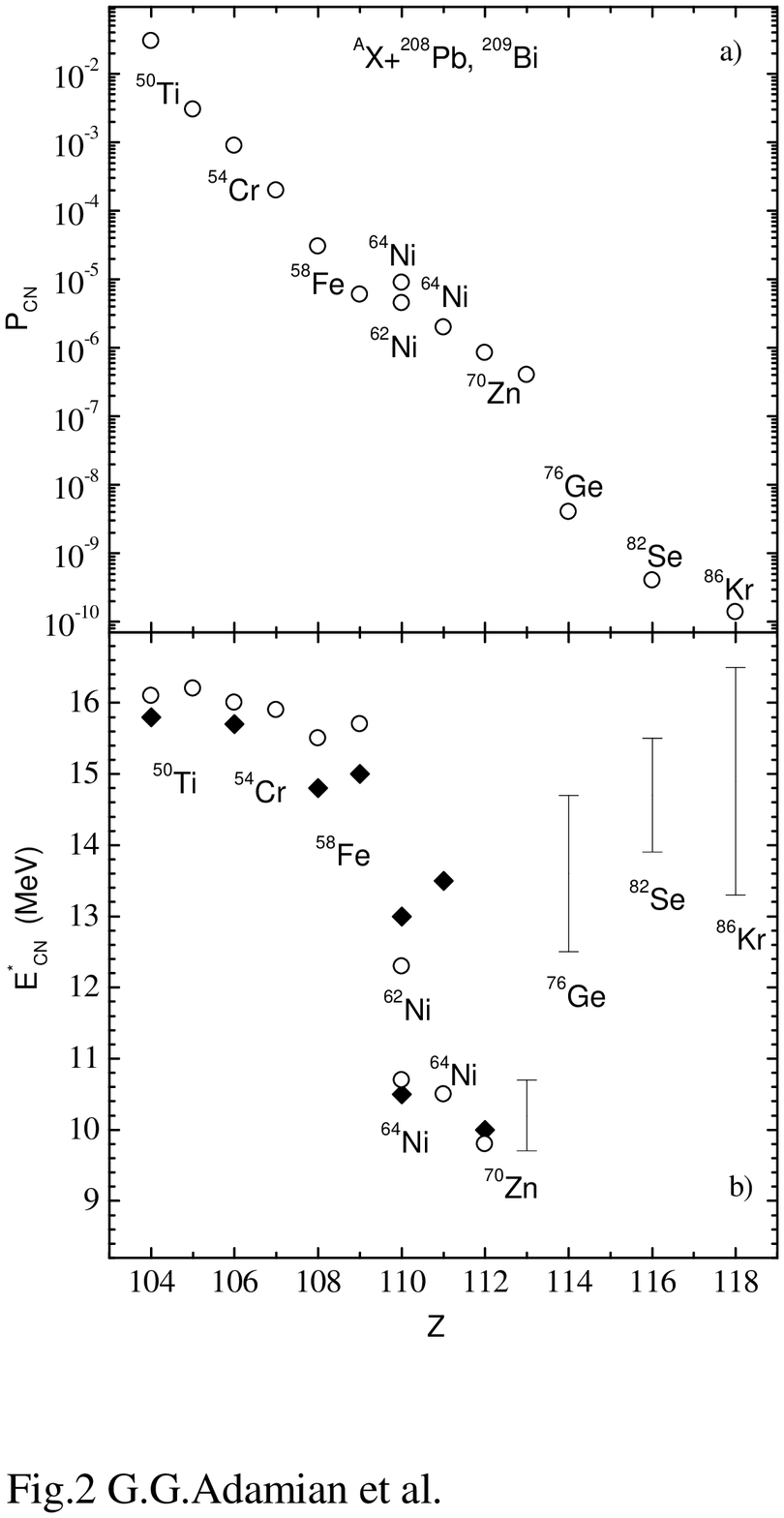,width=14cm}
\caption{a)
Calculated fusion probabilities $P_{CN}$ for cold fusion
in (HI,1n) reactions for the projectiles indicated
(open circles). The experimental data \protect\cite{SHOF,Muen}
are shown by solid diamonds.
For the compound nuclei with $Z$=104-112, the calculations were performed
with $Q$-values from Ref.~\protect\cite{Moel}.
b) Optimal excitation energies of the compound nuclei.
For the nuclei with $Z$=113,114,116 and 118,
the lower and upper limits of bars were calculated with $Q$-values
from Ref.~\protect\cite{Moel} and Ref.~\protect\cite{Moel1}, respectively.}
\label{2_fig}
\end{figure}

\begin{figure}
\psfig{figure=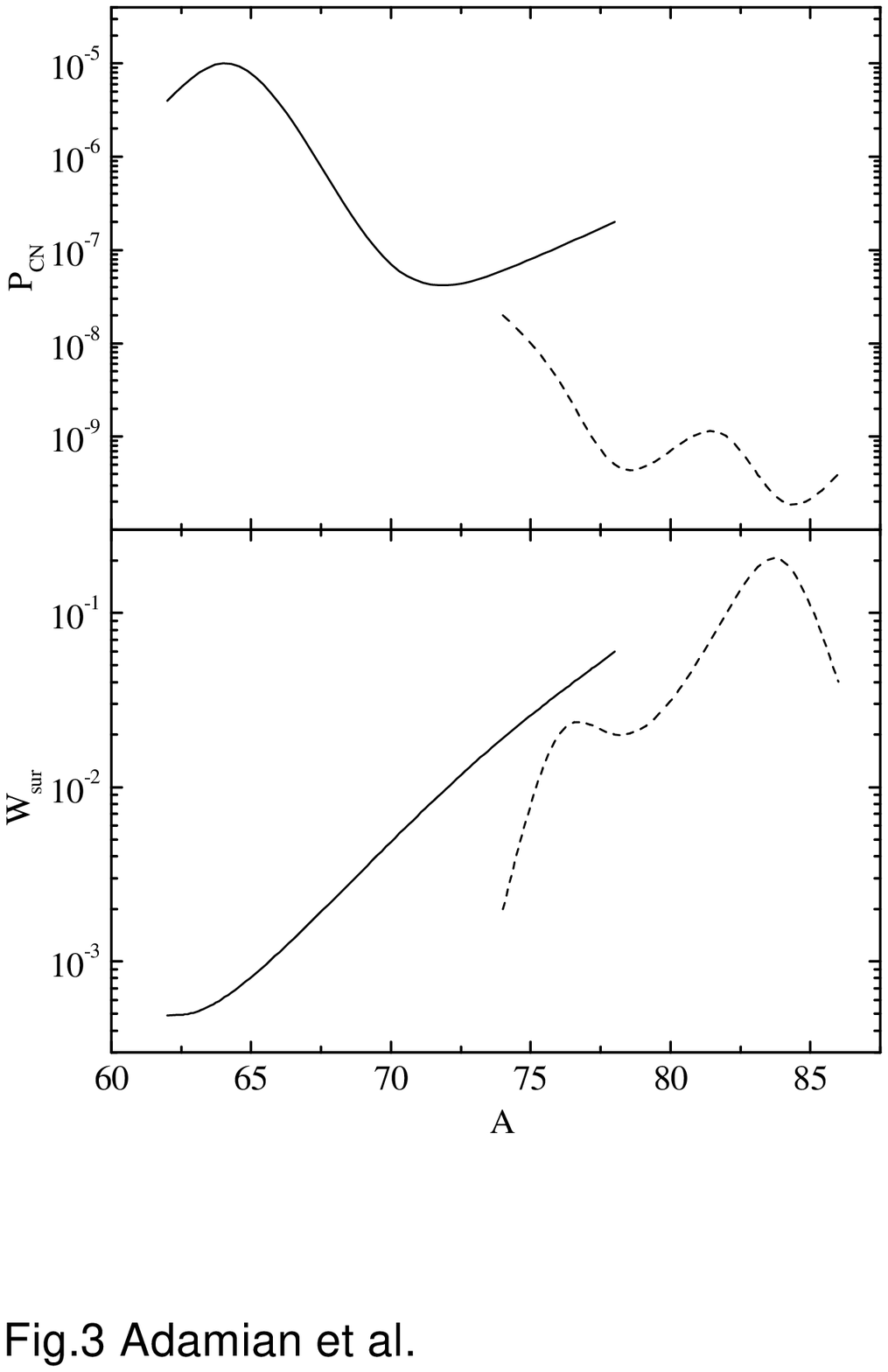,width=14cm}
\caption{The same as in Fig.~1  for the reactions
$^{A}$Ni+$^{208}$Pb (solid lines) and $^{A}$Ge+$^{208}$Pb (dashed lines).}
\label{3_fig}
\end{figure}

\end{document}